\documentclass[showpacs,floatfix,preprintnumbers,amsmath,amssymb]{revtex4}
\usepackage{graphicx}% Include figure files
\usepackage{dcolumn}% Align table columns on decimal point
\usepackage{bm}% bold math
\newcommand{\be}{\begin{equation}}
\newcommand{\ee}{\end{equation}}
\newcommand{\bea}{\begin{eqnarray}}
\newcommand{\beas}{\begin{eqnarray*}}
\newcommand{\eea}{\end{eqnarray}}
\newcommand{\eeas}{\end{eqnarray*}} 
\newcommand{\ba}{\begin{array}}
\newcommand{\ea}{\end{array}}
\begin{document}
%\magnification=1200
%\centerline{\bf An Optimum Interval Method for High Statistics}
%\vskip .3in

\title{Extending the Optimum Interval Method}
\author{S.~Yellin}
\affiliation{Department of Physics, Stanford University,
Stanford, CA 94305, USA}
\date{\today}
\begin{abstract}
The optimum interval method for finding an upper limit of a one-dimensionally
distributed signal
in the presence of an unknown background is extended to the
case of high statistics.  There is also some discussion of how
the method can be extended to the multiple dimensional case.
\end{abstract}
\maketitle

\section{Introduction}

One may compute the, say, 90\% confidence level ``Poisson statistics'' upper
limit of a signal by finding the value of the expected number of events
which would result in 90\% of random experiments having more events
in the entire experimental range than the number actually observed.
This method ignores the
distribution of the events.  A difference between the observed and
expected event distribution is a sign of background contaminating the
signal, in which case one should be able to get a stronger upper limit
by taking into account that difference. The
``optimum interval'' method~\cite{yellin} is a way of estimating an upper limit
in the presence of an unknown background.  The limit from this method is based
on selecting an interval within the experimental range which
contains especially few events compared with what would have been expected
from a true signal.  The method therefore avoids parts of the experimental
range with especially high background.  This interval is then used
to find an especially strong (low) upper limit, with the calculation taking
into account the method by which the interval was selected.  The result
is a true, though possibly conservative, frequentist confidence level.
But the method as published makes it practical only for
the case of low statistics.  With hundreds of events, it would
take too much computing time to generate the necessary tables.  For such a
case, this note proposes an alternative procedure for producing an upper limit
in the presence of unknown background.  Call this procedure the ``high statistics
optimum interval method'', while the originally published procedure will be
called the ``low statistics optimum interval method''.

Let's assume at first that events are characterized by some one-dimensional
value, $s$.  For example, $s$ might be the ``energy'' of the event.
Later the multidimensional case will be considered.  The set of events
consists of a ``signal'' as a function of $s$ plus an unknown
background.  There may also be a known background, but that can be considered
to be a part of the signal.  The ``size'' of a
proposed signal is characterized by the total number of events expected
from it in the experimental range of $s$.  If there is a known background,
the total number of events expected from it can then be subtracted from the
optimum interval upper limit
to give the size of the part of the signal which the experiment is designed
to measure.

More formally,
the total density distribution of the data within a one-dimensional
experimental range characterized by variable ``$s$'' is
the sum of the distribution from a non-negative signal, $\rho(s)$, plus an
unknown non-negative background, $\rho_U(s)$:
$\rho_T(s) = \rho(s) + \rho_U(s)$.
Take the experimental range to run from $s_a$ to $s_b$.
Instead of using $s$ to characterize events, make a change of variables to
$X(s)=\int_{s_a}^s \rho(s)$.  This new variable runs from $X(s_a)=0$ to
$X(s_b)=\mu$, the total number of signal events expected
within the experimental range.  Finding an upper limit
on the signal means finding an upper limit on $\mu$.
Call ``$x$'' the length of an interval in $X$, so that $x=X(s')-X(s)$ is
the expected number of events from the
signal between $s$ and $s'$.  With the published low statistics optimum
interval method, for each fixed number, $n$, one seeks the interval with
$n$ events which has the greatest $x$.
Of all intervals with $n$ events, the one with
largest $x$ is the optimum one for getting an upper limit in the
presence of background because it tends to be the one with
the lowest background relative to the signal.
The larger the observed value of maximum $x$ is for a given $n$, the stronger
is the rejection of the assumed signal.
For each $n$ a Monte Carlo generated table is
used to quantify how strongly the given signal is excluded.
The probability that the largest length with $\le n$ events is less than
$x$ is called ``$C_n(x;\mu)$''.  It is the tabulated measure of how strongly
the assumed signal is excluded by the data.  If $C_n(x;\mu)$, for
$x$ taken as the largest length in $X$ with $\le n$ events, is 90\%, then the
assumed signal is excluded to the 90\% confidence level.
The ``optimum'' $n$ is the
one that gives the strongest exclusion.  For this optimum $n$, the interval
which most strongly rejects the signal is the ``optimum interval''.
Monte Carlo is again used to generate a table used to
calculate the probability of the signal being excluded as
strongly as it was by the optimum interval.

Since $x$, $n$, and $\mu$ are all defined in a way that is invariant
under change of variables from the original $s$, this method is invariant
under a change of variables.  The method cannot be biased by how the
experimenter chooses to bin data because the data are not binned.  The
method is not very sensitive to cuts in the range of $s$ used to exclude
high backgrounds because it automatically avoids regions of high background
even if they are included in the range.  But the Monte Carlo calculation
of $C_n(x;\mu)$ can be time consuming for high $n$ and $\mu$.

\section{Extension to high statistics}

The following statements are equivalent:
``Largest length in $X$ with $\le n$ events is $< x$.'' $\equiv$
``No interval of length $x$ has $\le n$ events.'' $\equiv$
``The smallest number of events in intervals of length $x$ is $>n$.''
Thus $C_n(x;\mu)$ is also the probability that the
smallest number of events in intervals of length $x$ is $>n$.
One way of formulating a Monte Carlo method of computing 
this interpretation of $C_n(x;\mu)$
is to first define $F(s)=X(s)/\mu$, the cumulative probability
distribution of $s$ (the probability
that a random signal event will have a smaller measured
value than $s$).  Then $f=x/\mu=(X(s')-X(s))/\mu=F(s')-F(s)$ is the difference
in $F$ between two points.
Define  $y=(n-x)/\sqrt{x}$.
This definition was chosen so that for sufficiently large $x$,
$y$ from any particular fixed interval of length $x$ is approximately
distributed according to a standard normal frequency distribution -- a Gaussian
with mean zero and unit standard deviation.
Each Monte Carlo experiment takes the interval in $F$ from 0 to 1,
generates events uniformly in it with probability density $\mu$, and
finds the interval whose $f=x/\mu$ has the smallest $y$.
Call this smallest $y$ ``$y_{min}$''.  With a large number of
such Monte Carlo experiments one obtains the probability distribution
of $y_{min}$.  $C_n(x;\mu)$ is then the probability that $y_{min}
> (n-x)/\sqrt{x}.$

In the limit of very large $\mu$ and $x$, the result becomes
what we'll call ``$C_\infty(y_{min};f)$'', and it's independent of $\mu$.
Gaussian Brownian motion can be thought of as
the result of breaking a time interval into a huge number of equal small steps,
each of which adds an independent Gaussian random contribution with zero
mean and equal tiny variance.  In the limit of an infinite
number of infinitesimal steps, the resulting random path is
denoted by ``$w(t)$''.  With $w(0)$ initialized at zero, and the size of
the variances chosen to
give $w(t+1)-w(t)$ a standard normal frequency distribution, $w(t)$ is
called a ``standard Brownian process'' or a ``standard Wiener process''.
$C_\infty(y;f)$ is then the probability that
$$y_{min} = \min_{0\le t\le 1-f}\left[\frac{w(t+f)-w(t)}{\sqrt{f}}\right]$$
is greater than $y$.  It has been computed with a Monte Carlo program whose
technical details are described in Appendix A.  The function is
evaluated from a table whose interpolation in $f$ is simplified
by the empirical fact that if $C_\infty$ is tabulated in
$y'=y(1-0.3\, log(f)) - 1.7\, log(f)$ then the resulting function
of $y'$ varies slowly with $f$.  $C_\infty(y;f)$ decreases as $y$
increases with constant $f$, and it also decreases as $f$ decreases
with constant $y$.
%  $C_\infty(y;1)=0.5\,{\rm erfc}(y/\sqrt{2})$
%is an integral of the normal frequency distribution.

\begin{figure}
\includegraphics{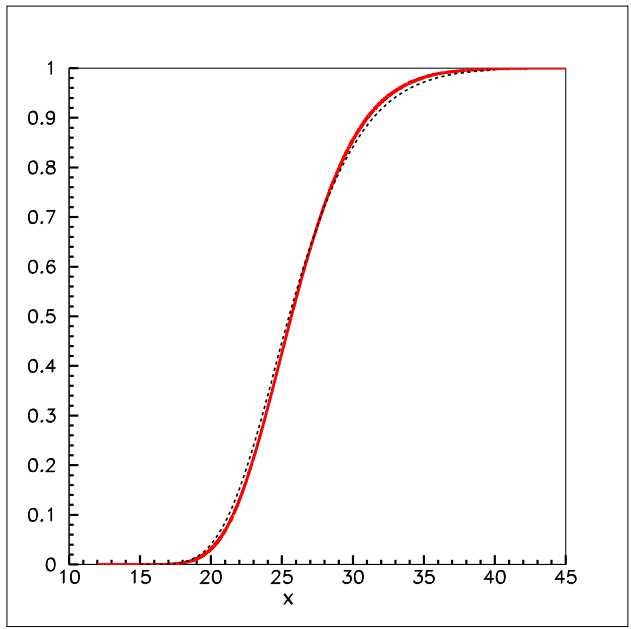}
\caption{Plots of $C_n(x;\mu)$ (solid) and $C_\infty(y;f)$ (dashed) for
$n=20$, $\mu=50$, $y=(n-x)/\sqrt{x}$, and $f=x/\mu$.}
\label{compareh}
\end{figure}

The definition of $C_\infty$ leads one to expect
$C_n(x;\mu) \approx C_\infty(y;f)$, where
$y=(n-x)/\sqrt{x}$ and $f=x/\mu$.  This approximation is only
valid for large $n$, but for large $x$, the probability of
small $n$ is negligeable.  Figure \ref{compareh} compares $C_{20}(x;50)$
with $C_\infty(y;f)$ for $y=(20-x)/\sqrt{x}$ and $f=x/50$.

\setkeys{Gin}{width=3.25 in} % All figures have width 3.25 inches.

If one finds $y_{min}$ for the data, $C_\infty(y_{min};f)$ says how strongly
the data reject the assumed signal as being too high.  One may then
vary $f$ in order to find the ``optimum'' one that gives the strongest
upper limit.  For a finite number of events in the data, call $x_m$
the largest gap between adjacent events.  For this interval, $y=-\sqrt{x_m}$.
Decreasing $f$ below $f_m\equiv x_m/\mu$ will give a larger value of $y$
with the smaller value of $f$, hence a smaller value of
$C_\infty(y;f)$; so $f<f_m$ cannot correspond to the ``optimum'' interval.
The search for the optimum interval over all $f$ is therefore equivalent
to a search restricted to $f>f_m$,
the $f$ corresponding to the largest gap between adjacent events.
One might prefer restricting $f$ to be greater than some even larger value,
$f_{min}$, if
on very small scales the event distribution is expected to be unreliable for
experimental reasons.
Or it may be that with a sufficiently large number of events, excessive
computation time is needed to find a good approximation to the minimum $y$
for all $f>f_m$.  So the following discussion
will assume $f$ is restricted to be greater than some $f_{min}$, which
includes the case $f_{min}=0$.

In principle, the optimum interval is optimized over
an infinite number of possible intervals.  But for any finite number of
events, only a finite number of intervals needs to be examined to find
the optimum one.
For a given number of events in the interval, it is the one with the
largest $f$, because $x=\mu\,f$ will then need the smallest $\mu$ to make
the expected number of events, $x$, be too large for the observed number
of events.
So the only intervals which need to be considered are those which begin just
after one event and end just before some other event.
First find for each $n$ the interval with $n$ events and with the largest $f$,
then compare
the computed upper limit $\mu$ for each $n$ and choose the smallest $\mu$.

\begin{figure}
\includegraphics{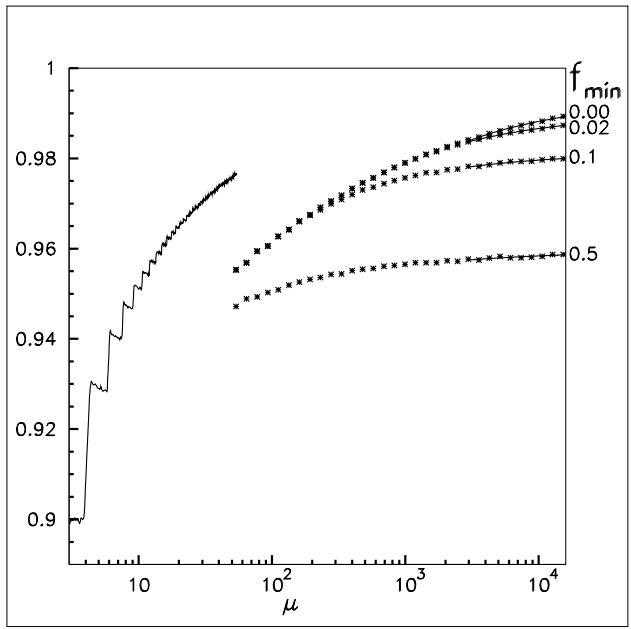}
\caption{90\% confidence level value, $\bar C_{Max}(0.9,f_{min},\mu)$
as a function of $\mu$ for $f_{min}=$ 0.00, 0.02, 0.1, and
0.5.  Points are generated by Monte Carlo.  The smooth curves at high $\mu$ are
fits of the form $A+B/\sqrt{\mu}$ to the points with $\mu>3500$.
For $\mu<54.5$, $\bar C_{Max}(0.9,0,\mu)$ is shown as calculated for the
low statistics optimum interval method.  While some of the
lack of smoothness of the low $\mu$ curve comes from statistical fluctuations
of the Monte Carlo program output, most of it represents the true behavior of
$\bar C_{Max}$, as is explained in Ref. \cite{yellin}.}

\label{CMaxbar}
\end{figure}

Call ``$C_{Max}$'' the maximum of $C_\infty(y;f)$ 
over all intervals with $f>f_{min}$.  A Monte Carlo program
can be used to compute the probability distribution of $C_{Max}$,
and thus can give the true confidence level by which the signal is
excluded.  Call $\bar C_{Max}(C,f_{min},\mu)$ the value of $C_{Max}$
for which the $C$ confidence level is reached for the assumed values
of $f_{min}$ and $\mu$.  Figure \ref{CMaxbar}
shows $\bar C_{Max}(0.9,f_{min},\mu)$ for various values of
$f_{min}$.  The lowest value of $\mu$ for which $\bar C_{Max}$ is
defined when the confidence level is 90\% is $\mu=2.3026$.  The highest
value of $\mu$ for which the $C_n$ of the low statistics method have
been tabulated is $\mu=54.5$.  Figure \ref{CMaxbar} shows results for the
low statistics
optimum interval when $2.3026\le\mu\le 54.5$, and for the high statistics
method when $\mu>54.5$.  $\bar C_{Max}$ has been tabulated only up to
$\mu=15310$, but can be extrapolated to $\mu>15310$ using the form
$\bar C_{Max}(C,f_{min},\mu)=A+B/\sqrt{\mu}$ with $A(C,f_{min})$
and $B(C,f_{min})$ fit to the calculated results with $\mu>3500$.
The extrapolation has been verified to
within the accuracy available from the number of Monte Carlo generated model
experiments at various values of $\mu$ up to 100000.

Since the computation of $y$ depends on the assumed value of $\mu$,
$C_{Max}$ is sensitive to $\mu$.  The optimum interval upper limit is
the signal size for which $\mu$, the total number of expected events in the
signal plus known background, satisfies $C_{Max} =
\bar C_{Max}(C,f_{min},\mu)$.  Methods for solving this equation are
described in Appendix B.

By having a choice of $f_{min}$, one is exposed to the
possibility of choosing $f_{min}$ so as to push the upper limit
in a desired direction.  Those using this method should choose
$f_{min}$ once at an early stage and stick with it.  In the absence
of a good reason to choose a non-zero value, $f_{min}=0$ is preferable,
because the smallest $f_{min}$ makes use of the most information from the data.

The low and high statistics optimum interval methods can be merged to
form what now will be called the ``optimum interval method''.
%  The highest
%value of $\mu$ for which tables have been generated for the $C_n$ of the
%low statistics method is $\mu=54.5$.
In both the computation of $C_{Max}$ and the computation of the table
for $\bar C_{Max}$, use $C_n$ for $\mu\le 54.5$, and use $C_\infty$ 
for $\mu>54.5$.  When the optimum interval method gives a result with
$\mu<54.5$, that result is the same as for the low statistics optimum
interval method; otherwise the result is the same as for the high
statistics optimum interval method.  In either case, the method for
computing $\bar C_{Max}$ means, for example, that  in the absence of
background, and independent of the true value of the signal, a 90\%
confidence level upper limit has a 90\% probability of being above
the true value.  This property of the method has been verified even near
the discontinuity in $\bar C_{Max}$ at $\mu=54.5$. And of course the
upper limit has a $>90\%$ probability of being above the true value if
there exists unknown background.

\section{Comparisons of Methods for Computing Upper Limits}

\begin{figure}
\includegraphics{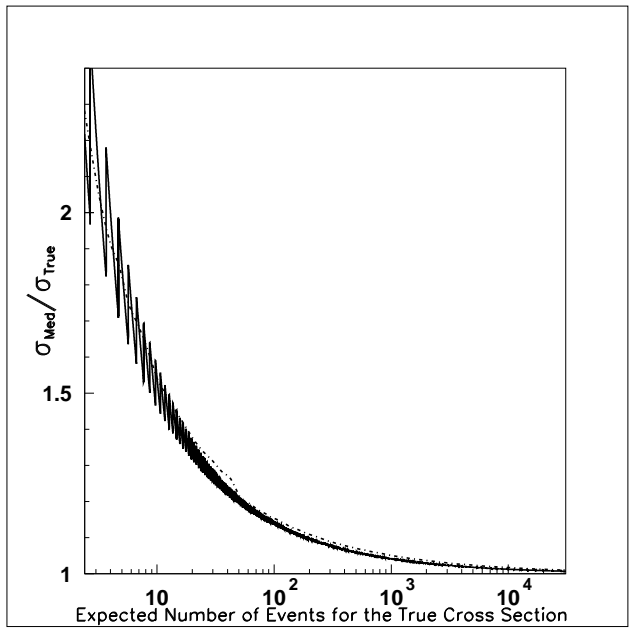}
\caption{Ratio between the median 90\% confidence level upper limit and the
true signal
as a function of $\mu$, the expected number of events in the true signal,
when there is no background.  The solid jagged line shows
the Poisson upper limit.  It is jagged because of the discrete nature of the
method -- for most $\mu$ it gives a stronger than 90\% upper limit.  The
dash-dotted line almost covered by the Poisson line is the 90\% confidence
level upper limit for $f_{min}=0$.
It is almost the same as the corresponding limit for $f_{min}=0.2$, which
is not shown.}

\label{noback}
\end{figure}

The optimum interval method can be compared with the Poisson
statistics upper limit.  For the case of no
background Fig. \ref{noback} shows the ratio between the upper limit
and the true signal for these two ways of computing upper
limits.  For this figure $\sigma_{Med}$ denotes the median value of the
computed upper limit signal size and $\sigma_{True}$ is the true signal
size.  The notation ``$\sigma$'' is used because this method was derived to
obtain an upper limit on a cross section, but in general $\sigma$ is
just something proportional to the total expected number of events
in the signal.

Both methods give approximately the same result in the absence of background.
As might be expected, when there is no background the Poisson upper limit
tends to be slightly stronger.  At those values of $\mu$ for which the
optimum interval method is
slightly stronger, the Poisson limit has a probability of less than 10\% of
finding an upper limit below the true signal size, while the optimum interval
method has exactly 10\% probability of making such a mistake.

\begin{figure}
\includegraphics{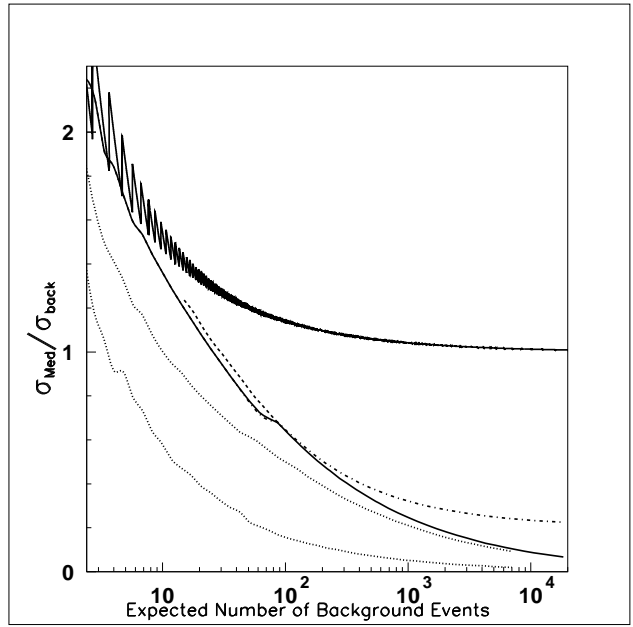}
\caption{Ratio between the median 90\% confidence level upper limit and the
background expected number of events
as a function of the expected number of events in the background
when there is no true signal.  The jagged solid line above
$\sigma_{Med}/\sigma_{back}=1$ is the Poisson upper limit.
The lower solid line is the optimum interval 90\%
confidence level upper limit for $f_{min}=0.0$, and the dash-dotted line
is for $f_{min}=0.2$.  The lower two dotted curves show the $f_{min}=0$
upper limit when half the background is known and when all of it is known.
The dashed segment just above the solid line
at fewer than 100 expected background events shows what the result would
be for $f_{min}=0.0$ if the pure high statistics optimum interval were used.}

\label{back}
\end{figure}

\begin{figure}
\includegraphics{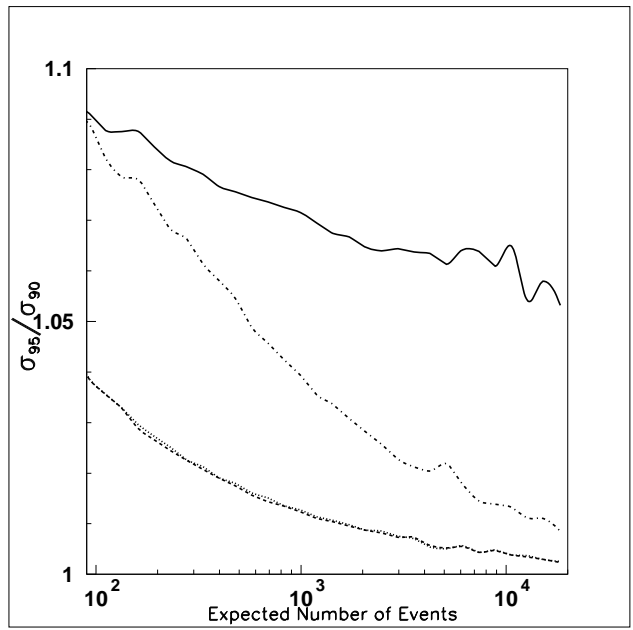}
\caption{Ratio between the median 95\% confidence level upper limit and the
median 90\% confidence level upper limit signal
as a function of the expected number of events.  For the case of the
background described in the text,
the solid curve is the ratio for $f_{min}=0.0$, and the dash-dotted one
below it is for $f_{min}=0.2$.  Below the curves for the background case
are two curves
almost on top of each other showing the zero background case: a dotted curve
for $f_{min}=0.2$ and a dashed curve for $f_{min}=0.0$.}
\label{test95vs90}
\end{figure}

As an example of what the method does with a background distributed
differently from the expected signal, consider a background that becomes
negligeable at one end of the experimental range.  To be more specific,
suppose the part of the signal whose upper limit is to be found has
density $\rho_S(s)$, and define
$X(s)=\int_{s_a}^s\rho_S(t) dt$.  With the change of variables from $s$ to $X$,
this part of the signal is distributed uniformly in $X$.  The example of
background used when making Fig. \ref{back} was, on the other hand, taken to
have density proportional to $X$, rather than uniform in $X$.
Fig. \ref{back} compares the Poisson 90\% confidence level
upper limit, the optimum interval limit with $f_{min}=0.2$,
and the optimum interval limit with $f_{min}=0.0$ for the case of
only such a background, which in all but the bottom two dotted curves
is assumed to be completely unknown.  In this case, the Poisson upper limit is
much worse than the optimum interval limits.  As might be expected,
for high statistics, $f_{min}=0.0$ (lower solid curve) gives a stronger upper
limit than $f_{min}=0.2$ (dot-dashed) because it's able to use a smaller optimum
interval where background is especially small near one end of the
experimental range.  The bottom two dotted curves show what happens when
half the background is considered known, so can be subtracted, and, for
the lowest curve, when all the background is considered known.  Appendix
B gives more details on how background is subtracted.
Using the high statistics optimum interval method
with its Gaussian approximation, even when $\mu<54.5$, gives a result (dashed)
slightly weaker than is given by the optimum interval method with the low
statistics method for $\mu<54.5$.

Because the computation of $\bar C_{Max}$ for $\mu>15310$ was done with a fit
to the form $A+B/\sqrt{\mu}$, one might worry that inaccuracies in the
extrapolation will make results inaccurate.  The effect of an inaccurate
value for $\bar C_{Max}$ is to make the true confidence level somewhat
different from the intended one.  The probability of mistakenly
getting a ``90\% confidence level'' upper limit below the true signal
should be exactly 10\% when there's no unknown background.  The Monte
Carlo calculation used to make
Fig. \ref{noback} used 50000 model experiments for $\mu<2500$, 5000 model
experiments for $2500<\mu<20000$, and 500
for $\mu=100000$.  As would be expected for a correct computation of
$\bar C_{Max}$, mistakes occurred in $10$\% of the model experiments to
within the statistical errors of the limited number of trials.
Similar tests have been done for
95\% and 99.5\% confidence levels up to $\mu\approx 20000$.  Furthermore,
upper limits for high $\mu$ are not very sensitive to the
choice of value for the confidence level.  Figure \ref{test95vs90} shows
examples of the ratio between the 95\% confidence level upper limit and the
90\% confidence level upper limit. Curves are irregular largely because of
the limited number of Monte Carlo experiments used in their calculation.

\section{Extension to even higher statistics}

The method described here may be extended to extremely
high numbers of events.  Today's desktop computers can compute the
optimum interval upper
limit for 10000 events in seconds.  But since the computing time
grows like the square of the number of events, with enough
events the method becomes impractical.  There's a way to approximate
the optimum interval method in a way that instead grows linearly with
the number of events.  Recall that in Sections I and II evnts were
characterized by some one-dimensional
parameter, $s$, such as energy, with $F(s)$ the fraction of events expected
to be below $s$ for the assumed event distribution without background.
Bin the data in 1000 or more equal bins in $F$.  The computer time needed
to do the binning grows linearly
with the number of events.  Then consider only those intervals which consist
of one or more consecutive bins.  The computer time for finding the optimum
such interval is independent of the number of events.  The larger the number
of bins, the closer is this method to the optimum interval method without
binning.  At this time, tables and software for this modification of the
optimum interval method have not been produced.  If they are produced,
they will be for a particular choice of the number of bins of $F$.  That
choice can easily be independent of what the experimenters want their
result to be; so choice of binning would
not be a way experimenters could inadvertently bias results.

\section{Extension to multiple dimensions}

Some experimentalists need a generalization
of the optimum interval method to more than one dimension.  The impetus for
producing this section and the corresponding Appendix C was a
discussion\cite{Fisher} with physicists who were already planning to write
software extending the low statistics optimum interval method to two
dimensions.  Some ideas are discussed here concerning the extension to an
arbitrary number of dimensions and to the case of high statistics, but
these ideas have not yet been implemented in software.

Suppose instead of data described in terms of a distribution in one-parameter,
$s$, there are $D$ dimensions, $s_1,\ s_2,\, ...,\ s_D,$ with the signal
distributed according to some density function $\rho_D(s_1,...,s_D)$.
Appendix C shows how to transform coordinates from $(s_1,...,s_D) \rightarrow 
(r_1,...,r_D)$ in a way which maps $\rho_D$ into a uniform distribution
within the unit ``generalized cube'' with $0< r_p < 1$ for all
$1\le p\le D$.  As in the case of one dimension, call ``$\mu$'' the expected
number of events in the entire range of the experiment, which in terms
of coordinates $\vec r$ is the unit generalized cube.  Any region
within the unit generalized cube occupies fraction $f$ of the generalized
cube, and $f$ is also the generalized volume of the region.  The expected
number of events within a region inside the unit generalized cube is
$x=\mu f$.

As in the case of one dimension, selection of the optimum region for
an upper limit on $\rho_D$ begins by choosing some set of allowable
regions within the total experimental range.  In the case of one
dimension, the allowable regions are intervals.  For
each $n$ find the allowable region with $\le n$ events having the largest 
expected number of events, $x$.  If $x$ is too large for the observed
$n$, that means the assumed signal, $\rho_D$, is excluded by observation as
being too large.  For $D$ dimensions take the allowable regions to be
``GRPs'', generalized rectangular parallelepipeds with all edges parallel
to coordinate axes in $\vec r$ space.  Explicitly, a GRP is a region of
$\vec r$ space
defined with some fixed set of $a_p < b_p$ as the set of $\vec r$ for which
$a_p < r_p < b_p$ for all $1\le p \le D$.  It's a line segment in one
dimension, a rectangle in two dimensions, and
a rectangular parallelepiped in three dimensions.  The generalized volume of a
GRP is
$$ f = \prod_{p=1}^D (b_p-a_p).$$
``$C_{nD}(x;\mu)$'' is defined as the probability for $D$ dimensions that the
largest GRP with $\le n$ events would have generalized volume less than
$f=x/\mu$ if the assumed $\rho_D$ were correct.  For $D=1$ this
definition is equivalent to the one for $C_n(x;\mu)$.  As in the case of
$D=1$, $C_{nD}(x;\mu)$ is a measure of how strongly the assumed $\rho_D$
is excluded by the data.  Appendix C discusses how to evaluate $C_{nD}$.

The optimum GRP, the generalization of the optimum interval of one dimension,
is the one for which $x$ and $n$ give the largest value of $C_{nD}(x,\mu)$,
which one may call ``$C_{Max}$''.  Define
``$\bar C_{Max,D}(C,\mu)$'' as the value such that a fraction $C$ of random
experiments with a model having that $\mu$ (and no unknown background)
will give $C_{Max}<\bar C_{Max,D}(C,\mu)$.  $C$ is the confidence level
by which the model is excluded because the data show too small a signal.

$C_\infty$ of one dimension can be generalized to
$C_{\infty D}$.  First notice that $C_{nD}(x;\mu)$ is also the probability
that the smallest
number of events in regions with expected number of events, $x$, is $>n$.  As
for the case of $D=1$, define $y=(n-x)/\sqrt{x}$.  If the smallest $y$ in GRPs
with generalized volume $f$ is $y_{min}$, then $C_{nD}(x;\mu)$ is the
probability that $y_{min}>(n-x)/\sqrt{x}$.  In the limit of large $\mu$
and $x$, the probability distribution of $y$ in any single GRP is Gaussian
with zero mean and unit standard deviation.  Now generalize standard
Brownian motion to $D$ ``time'' dimensions: Break the entire unit generalized
cube into infinitesimal pieces, each of which contributes to any region
containing the piece an independent signal distributed according to a Gaussian
with zero mean and variance such that the entire unit generalized cube has
a signal of unit variance.  The standard deviation of the signal in a
region with generalized volume $f$ is $\sqrt{f}$.  Thus in any region with
generalized volume $f$, ${\rm Signal}/\sqrt{f}$ has zero mean and unit
standard deviation, exactly as does $y=(n-x)/\sqrt{x}$ in the limit of
high statistics.
Define ``$C_{\infty D}(y;f)$'' to be the probability that for all
GRPs of generalized volume $f$ within the unit generalized cube, the
${\rm Signal}/\sqrt{f}$ is greater than $y$.  Appendix C discusses its
computation.  For $D=1$ its definition is the same as that of $C_\infty$
of section II.

Finally, $\bar C_{Max}(C,f_{min},\mu)$ can be generalized to 
$\bar C_{Max,D}(C,f_{min},\mu)$.  If only $f_{min}=0$ is considered,
the same function might as well be written as ``$\bar C_{Max,D}(C,\mu)$''.
It is defined in the
same way as the function written the same way for the low statistics case.
The difference is only that the low statistics function applies only
for low $\mu$, while the high statistics function applies for high $\mu$.
$\bar C_{Max}(C,\mu)$ can be computed with a Monte Carlo program for
low $\mu$ once the $C_{nD}$ functions are available for low $\mu$, and
for high $\mu$ once the $C_{\infty D}$ function is available.

\section{Conclusions}

The optimum interval method has been extended to the case of high
statistics by making a Gaussian approximation to the probability distribution
of events in each subinterval of the experimental range.  Even with this
approximation, the method is a true, though
possibly conservative, frequentist 
upper limit, with the probability of mistakenly getting too small a result
being at most one minus the confidence level.

Software and tables are available\cite{software} for applying this method to
actual data.  Once 1000 events have been manipulated
into a form appropriate for the software, it takes a 730 MHz Pentium III
computer about 0.022 seconds to compute the upper limit.
Computation time is approximately proportional to the square of the number
of events.

Although the extension of the optimum interval method to multiple dimensions
can be computationally very intensive, and
has not yet been implemented in software, ideas for how it should be done
have been presented.

\appendix
\section{Technical details of the computation of the relevant functions}

In the computation of $C_\infty$, each Monte Carlo trial begins
by breaking the interval (0,1) into a very large number, $N$, of
subintervals.  For each subinterval generate an independent random
value according to a standard normal frequency distribution.
Restrict oneself to lengths $f$ within (0,1) which are integer multiples
of $1/N$.  For each such $f$ for which $C_\infty(y_{min};f)$ is to be
computed, allow it at first to only be at positions within (0,1) such
that its endpoints are on endpoints of subintervals.  Such a length $f$
interval contains $Nf$ subintervals.  For any such length
$f$ interval, scale the contribution from all subintervals it contains
by a factor of $1/\sqrt{Nf}$ to produce a total signal $y$ with zero mean
and unit standard deviation.  Move the
interval of length $f$ through (0,1) in steps of size $1/N$,
searching for $y_{min}$, the smallest $y$.  That's the result of
this one Monte Carlo trial.  Do a huge number of them, and tabulate
the distribution of $y_{min}$ for the various values of $f$.

With finite $N$, the computed estimate of $C_\infty$ is systematically
shifted above its true value because the minimum signal
occurs somewhere inside one of the tiny subintervals, not at an end of
one as was assumed in the proposed Monte Carlo method.
The size of this systematic shift is of order the standard deviation
of the signal contribution to subintervals.  I.e., it's of order
$1/\sqrt{N}$.  Thus the resulting distribution of $y_{min}$ will be
systematically shifted too high by an amount of order $1/\sqrt{N}$,
and the result will not get arbitrarily accurate as the number of Monte
Carlo experiments grows arbitrarily high. I expect the optimum time for a
given accuracy would be achieved by choosing the number of Monte experiments
to be proportional
to $N$, so that both the systematic error from finite size of the subintervals
and the statistical Monte Carlo error would decline like $1/\sqrt(N)$.
The computing time then would grow like
$1/{\rm accuracy}^4$, instead of the less rapid growth $1/{\rm accuracy}^2$
usually characteristic of Monte Carlo calculations.
 
To restore the Monte Carlo error to a $1/\sqrt{N}$ decline with $N$, correct
for the systematic shift from finite $N$ by including a Monte Carlo
estimate of what the minimum $y$ would be within each subinterval.
I.e., pick a random value
according to the probability distribution of the minimum
within the subinterval.  Call the endpoints of a subinterval
``$t_0$'' and ``$t_1$'', with $t_1-t_0=1/N$.
In the evaluation of $C_\infty$ the Monte Carlo program finds
$y(t)=[w(t+f)-w(t)]/\sqrt{f}$ at $t$ equal to $t_0$ and $t_1$.  We want
the distribution of what would be found if the minimization were also
done over all points $t_0<t<t_1$.  To do this note that as the
interval slides from $t=t_0$ to $t=t_1$ it loses infinitesimal
random Gaussian contributions from its low end and gains them at its high
end.  Thus the change in $y$ as $t$ increases
is itself a Gaussian Brownian process.  When $t$ increases by $1/N$,
$[w(t+f)-w(t)]$ loses a random contribution with variance $1/N$ from
its low end and gains an independent random contribution with variance
$1/N$ at its high end, for a total variance of $2/N$.  As $t$ increases
from $t_0$ to $t_1$, 
$y=[w(t+f)-w(t)]/\sqrt{f}$ changes by a random value with variance
$\sigma^2=2/(Nf)$.
Form $x$ from $t$ by a shift and
rescaling to put $x=0$ at $t=t_0$ and $x=1$ at $t=t_1$.  The change
in $y$ is proportional to that from a standard
Brownian motion rescaled by a factor which makes the total variance
for the change from $x=0$ to $x=1$ be $\sigma^2$ instead of unity.  We may
therefore write 
\be
y(t)=\frac{w(t+f)-w(t)}{\sqrt{f}} = \frac{w(t_0+f)-w(t_0)}{\sqrt{f}}
 + \sigma \tilde w(x),
\label{tildew}\ee
where $\tilde w(x)$ is a standard Brownian process with $v=\tilde w(1)$
constrained to give the correct value of $y(t_1)$.
To find the minimum of $y(t)$ as $t$ varies between
$t_0$ and $t_1$, find the minimum of $\tilde w(x)$ as $x$ runs from 0
to 1 subject to the constraint that $\tilde w(1)=v$.  
%To do this, define
%\be \begin{array}{cl}
%\tilde w(x) &= \sqrt{\frac{N}{2}}\Big([w(t_0 + x/N + f) - w(t_0 + x/N)] -
%[w(t_0+f) - w(t_0)]\Big)\\
%            &= \sqrt{\frac{N}{2}}\Big([w(t_0 + f + x/N) - w(t_0 + f)] -
%[w(t_0+x/N) - w(t_0)]\Big).\\
%\end{array}\label{tildew} \ee
%The form after the first ``$=$'' sign in Eq.\ref{tildew} is convenient
%for seeing that when
%$z$ is the minimum of $\tilde w(x)$ for $0\le x\le 1$, then the
%minimum value of $[w(t+f)-w(t)]/\sqrt{f}$ within $t_0\le t\le t_1$ is
%$[w(t_0+f)-w(t_0)]/\sqrt{f} + z\sqrt{2/fN}$.  The form after the second
%``$=$'' sign in Eq.\ref{tildew} is convenient for seeing that for
%$0\le x/N \le f$, $\tilde w(x)$ is a standard Brownian process: the first pair
%of terms within square brackets can be thought of as a sum of a huge number
%of independent Gaussian distributed contributions from $t_0+f$ to $t_0+f+x/N$,
%and the second pair can be considered a sum of contributions from
%$t_0$ to $t_0+x/N$.  So long as $x/N<f$, all these contributions
%are independent of each other, and their number is proportional
%to $x$.  Take $N$ large enough so that $1/N<f$.  We have $\tilde w(0)=0$,
%and the factor of $\sqrt{N/2}$ was chosen to make the variance of
%$\tilde w(1)$ unity.  So $\tilde w(x)$ is a standard Brownian process.
For the purpose of correcting the minimum in the Monte Carlo program with
finite N, the minimum of $\tilde w(x)$ over the range $0\le x\le 1$
was chosen randomly from its probability distribution given
fixed $v=\tilde w(1)$.  This probability distribution will now be derived.

The minimum of $\tilde w(x)$ for $0\le x\le 1$ can be no larger than the
minimum of 0 and $v$, because $\tilde w$ is equal to each of those values
at the endpoints of its range.  Consider only 
$z\le min(0,v)$.  For a given such $z$ the minimum of
$\tilde w(x)$ for $0<x<1$ is $\le z$ if and only if somewhere between
$x=0$ and $x=1$ $\tilde w(x)$ crosses $z$.  
Call ``$x_z$'' the value of $x$ for which $\tilde w(x)$ first
crosses $z$.  Define $w_R(x)$ to be the same as $\tilde w(x)$ when $x\le x_z$,
but for $x > x_z$ reverse the sign of each infinitesimal Gaussian contribution.
The new function is reflected through $z$ for $x>x_z$.
Since reversing the sign of a Gaussian contribution with zero mean leaves it
a Gaussian contribution with zero mean, $w_R(x)$ is also a standard Brownian
process.  For $x\ge x_z$, $w_R(x)=z-(w(x)-z) = 2z-\tilde w(x)$, and its
first crossing of $z$ is at the same $x_z$ as $\tilde w$.  So any
$\tilde w(x)$ for which the minimum is $< z$ and for which $\tilde w(1)=v$
defines a $w_R(x)$ for which $w_R(1)=2z-v$.  Since $w_R(1)=2z-v<z<0=w_R(0)$,
any $w_R(x)$ for which $w_R(1)=2z-v$ must cross $w_R(x)=z$.  Call the first
such value of $x$ ``$x_z$''.
The reflection of $w_R$ about $z$ for $x>x_z$ then
defines a $\tilde w(x)$ whose minimum is below $z$.
There is, therefore, a one-to-one correspondence between a) standard Brownian
processes whose minimum between $x=0$ and $x=1$ is below $z<min(0,v)$ and
for which $\tilde w(1)=v$ and b) standard Brownian processes for which
$w_R(1)=2z-v$.  The probability distribution of $w_R(1)$ is the normal
frequency distribution.  So the probability
of the minimum of $\tilde w(x)$ being
below $z$ in the range $0<x<1$ for $\tilde w(1)$ within some tiny
$\delta$ of $v$ is proportional to $\delta\, exp(-(2z-v)^2/2)$.
The probability of $\tilde w(1)$ being within the same $\delta$ of
$v$ is proportional to $\delta\, exp(-v^2/2)$, with the same
normalizing factor.
%Then for $z\le min(0,v)$
%the following statements are equivalent:
%``Minimum of $\tilde w(x) \le z$ and $\tilde w(1)$ is within a tiny $\delta$
%of $v$'' $\equiv$ ``Minimum of $w_R(x) \le z$ and $w_R(1)$ is within a tiny
%$\delta$ of $2z-v$'' $\equiv$ ``$w_R(1)$ is within a tiny $\delta$
%of $2z-v$''. 
Since the probability of A given B is equal to the probability of (A and B),
divided by the probability of B, we have the probability that the minimum of
$\tilde w(x)$ is $\le z$ given $\tilde w(1)=v$ is
\be
P = \frac{e^{-(2z-v)^2/2}}{e^{-v^2/2}} = e^{2z(v-z)}.\label{minw}
\ee
Various mathematical references\cite{lalley,gikhman} on stochastic processes
give similar, but more rigorous, derivations of equations related to \ref{minw}.
Random $z$ will have the desired probability distribution if one
first chooses random $P$ uniformly over (0,1), then solves the equation
relating $P$ and $z$ for $z$: $z=\left(v-\sqrt{v^2-2\,ln(P)}\right)/2.$
Replace $\tilde w(x)$ with $z$ in Eq. \ref{tildew} to get a Monte Carlo
minimum $y(t)$ over the range $t_0<t<t_1$.
Although this method eliminates most of the systematic error which would
otherwise be caused by the finite value of $N$,
large $N$ is still desireable because assuming independently random $z$ for
different subintervals ignores some correlations.

\section{Solving $C_{Max}=\bar C_{Max}$}

The data for which the equation $C_{Max}=\bar C_{Max}$ is to be solved
can be expressed as a set of values of the cumulative probability, $F=X/\mu$,
as introduced in section II.  Assume the signal can be expressed as a sum
of a part for which one wants an upper limit plus a part consisting of a
known background.  The total expected number of events below $s$ is
$X(s)=X_S(s)+X_B(s)$; $X_S(s)$ is the expected number below $s$ from the
part of the signal whose upper limit is to be determined, and $X_B(s)$
is the expected number below $s$ from the known background.  
$X_S(s_b)\equiv \mu_S$, $X_B(s_b)\equiv\mu_B$, and $\mu=\mu_S+\mu_B$.
For the known background, $\mu_B$ is assumed to be known.
The set of $F$ for the events can be computed for any trial value of $\mu$
from $F=(1-\mu_B/\mu)F_S + (\mu_B/\mu)F_B$, where $F_S=X_S/\mu_S$ and
$F_B=X_B/\mu_B$.  From the trial value of $\mu$
and from the resulting set of $F$ one may find $C_{Max}$, and one may
compare it with $\bar C_{Max}(C,f_{min},\mu)$.  Once a $\mu$ has been found
which makes the two equal, subtraction of the known background is
done by converting the upper limit on $\mu$ into one on $\mu_S=\mu-\mu_B$.

The equation
$C_{Max} = \bar C_{Max}(C,f_{min},\mu)$ for the $\mu$ corresponding
to the upper limit can be solved using CERNLIB\cite{CERNLIB} routine
RZERO, which finds the zero of a function of one real variable.  This
routine can also be used for the upper limit in the case of high
statistics.  But for high $\mu$, and with no background subtraction,
there's another method which usually converges faster.

The optimum interval is the one which requires the lowest upper limit $\mu$
to be excluded at confidence level $C_\infty(y;f)$.
For the optimum interval with $f>f_{min}$, $C_\infty(y;f) =
\bar C_{Max}(C,f_{min},\mu)$ for the upper limit value of $\mu$.
The inverse function of $\bar C_{Max} = C_\infty(y;f)$ is
$y=y_\infty(\bar C_{Max};f)$.  From
$y=(n-x)/\sqrt{x}=(n-\mu\, f)/\sqrt{\mu\, f}$
one may then solve for $\mu$.  The interval with $f>f_{min}$ with the
lowest $\mu$ is the optimum interval, and $\mu_F=\mu-\mu_B$ is the
``high statistics optimum interval method'' upper limit
which we seek.  Since $\bar C_{Max}$ does have some dependence on $\mu$,
an iterative procedure is needed: guess an initial $\mu$ and find its
corresponding $\bar C_{Max}$.  Then compute the
upper limit $\mu$, and take this improved estimate for $\mu$ to get
an improved estimate for $\bar C_{Max}$ in the next iteration.

The function $C_\infty(y;f)$ was computed with a Monte Carlo program only
for $0.01 < f < 1.0$.  Its inverse, $y_\infty(C;f)$, is also only tabulated
for $f$ in that range.  But for a sufficiently large number of events in
the experiment, the high statistics optimum interval method requires
evaluating $y_\infty(C;f)$ for $f<0.01$.

For finding the minimum $y$ of all intervals with expected fraction $f$
of the events, continuously move the interval of
size $f$ along it, while seeing if the current $y$ is the lowest.  
$C_\infty(y_\infty;f)$ is the probability that the lowest $y$ is
greater than $y_\infty$.  Now
imagine the entire experimental range broken into $p$ pieces.  The intervals
which are fractions $f$ of the whole range are fraction $f_0=pf$ of each
piece.
Each of the $p$ pieces has probability $C_\infty(y_\infty;f_0)$ of having
its lowest $y$ greater than $y_\infty$.  The probability of all the $p$
pieces having lowest $y$ greater than $y_\infty$ is $C^p_\infty(y_\infty;f_0)$.
These considerations lead to the approximation
\be C_\infty(y;f)\approx C^p_\infty(y;f_0),\label{Capprox}\ee
with $p=f_0/f$.  This approximation for $C_\infty(y;f)$
is not exact because the measurement of  minimum $y$ separately in each
piece misses those size $f$ intervals which overlap two adjacent
pieces.  The approximation can be improved by decreasing the effective number
of $f$ length intervals in the whole region, $1/f$, by a little bit for each
of the $p-1$ boundaries between pieces.  I.e, the approximation will be
better if $1/f = p(1/f_0) - s(p-1)$, for some $s$ of order unity.  Solving for
$p$ gives
\be p = \frac{1/f - s}{1/f_0 -s}.\label{pest}\ee
Solving both sides of Eq. \ref{Capprox} for $y$ gives
\be y_\infty(C;f) \approx y_\infty(C^{1/p};f_0).\label{yapprox}\ee
 
Empirically, $s=0.94$ works fairly well, according to tests with relatively
large $f_0$ for $0.01 < f < 1$, but for the smaller values of $f_0$, it's
hard to distinguish between various choices of order unity for $s$.

Although Eqs. \ref{Capprox}, \ref{pest}, and \ref{yapprox} were motivated
by choosing $p$ to be an integer number of pieces, they can be generalized
to non-integer $p$.  To see this, suppose that not only
does $f_0=pf$ for integer $p$, but $f_1=qf$ for another integer, $q$.  Then
Eq. \ref{Capprox} implies $C_\infty(y;f)\approx C^p_\infty(y;f_1)$ which,
along with Eq. \ref{Capprox}, gives for integer $p$ and $q$
$C_\infty(y;f_1)\approx C^{p/q}_\infty(y;f_0)$, where
$$p/q = \frac{(1/f - s)/(1/f_0 -s)}{(1/f-s)/(1/f_1-s)} =
\frac{1/f_1 -s}{1/f_0 -s}.$$
These equations are equivalent to
Eqs. \ref{Capprox} and \ref{pest}, with integer $p\rightarrow$ fraction $p/q$. 

The extrapolation of Eq. \ref{yapprox} from $f_0$ to $f$ may be poor if $f$
is too small, because $C^{1/p}$ may be too close
to unity for the table used by $y_\infty$ to give an adequate
approximation.  In such a case one may fall back to using $f_0=1$, for which
$C_\infty$ and $y_\infty$ can be calculated without resorting to Monte
Carlo generated tables.  The probability distribution of $y$ for
$f=1$ is the normal frequency distribution,
for which publicly available programs can compute
the integral and the integral's inverse.
Examples are DFREQ and DGAUSN of CERNLIB\cite{CERNLIB}.  For $f_0=1$
the accuracy for low $f$ is somewhat improved by using $p=(1/f - 0.946)/0.051$
instead of Eq. \ref{pest} with $s=0.94$.  This modified form for $p$ is
used for $f_0=1$ in table~\ref{yextrap}, which shows that the
extrapolation is relatively independent of the value of $f_0$ used.  
The Monte Carlo programs used to generate the tables used by
software implementing the high statistics optimum interval method
often had to extrapolate from $f_0=0.01$, but have so far never required
extrapolation from $f_0=1$.

\begin{table}
\caption{Comparison of the approximation of Eq. \ref{yapprox} for
various values of $f_0$.}
\label{yextrap}
\begin{tabular}{|r|r||c|c|c|c|}
\hline
f      & C  & $f_0=0.01$ & $f_0=0.02$ & $f_0=0.04$ & $f_0=1.0$ \\
\hline
0.0005 & 0.93 & -4.66 & -4.65 & -4.65 & -4.63 \\
       & 0.96 & -4.79 & -4.77 & -4.77 & -4.75 \\
       & 0.99 & -5.09 & -5.09 & -5.12 & -5.02 \\
\hline
0.0020 & 0.93 & -4.33 & -4.34 & -4.34 & -4.33 \\
       & 0.96 & -4.47 & -4.48 & -4.47 & -4.46 \\
       & 0.99 & -4.79 & -4.77 & -4.77 & -4.75 \\
\hline
0.0040 & 0.93 & -4.16 & -4.16 & -4.16 & -4.18 \\
       & 0.96 & -4.30 & -4.31 & -4.30 & -4.30 \\
       & 0.99 & -4.63 & -4.63 & -4.63 & -4.61 \\
\hline
0.0100 & 0.93 & -3.92 & -3.91 & -3.91 & -3.96 \\
       & 0.96 & -4.07 & -4.07 & -4.07 & -4.10 \\
       & 0.99 & -4.42 & -4.42 & -4.42 & -4.41 \\
\hline
\end{tabular}
\end{table}

\section{Technical details for the extension to multiple dimensions}

The optimum interval method in multiple dimensions begins with a
transformation of coordinates from initial coordinates, $s_1,...,s_D$,
with signal density $\rho_D(\vec s)$, into
ones for which the entire experimental range is in a unit generalized cube
with uniform transformed density.
Define $\rho_p$ for $0\le p\le D-1$ inductively by
\be
\rho_{p-1}(s_1,...,s_{p-1}) = \int_{-\infty}^{+\infty} dt
\rho_p(s_1,...,s_{p-1},t),
\ee
and then define $r_p$ for $1\le p\le D$ by
\be
r_p(s_1,...,s_p) = \frac{1}{\rho_{p-1}}\int_{-\infty}^{s_p} dt
\rho_p(s_1,...,s_{p-1},t). \label{stof}
\ee
The expected number of events in the entire experiment is $\mu=\rho_0$.

In order for a Monte Carlo program to compute $C_{nD}(x,\mu)$, and in order
to evaluate the optimum generalized rectangular parallelepiped (GRP)
for the actual data,
it is necessary to find for each number of events, $n$, the maximum
generalized volume of a GRP for Monte Carlo or real data.
To be a candidate for a maximal generalized volume GRP for a given $n$, each
generalized face
must butt up against one of the event points; otherwise $n$ could be kept
constant while expanding the GRP
along the normal to that generalized face until an event point is encountered.

An algorithm is needed for the computer program to find all GRPs which could
have the maximal generalized volume for each $n$.  For this purpose,
assume there are $M$ events ordered so that $s_1(i)$
increases with $i$ as $i$ runs from 1 to $M$.  For any non-negative $\rho_D$,
Eq.\ref{stof} implies that if the $s_1(i)$ are in increasing order, so are the
$r_1(i)$.  To simplify allowing boundaries of the unit generalized cube
to also be GRP boundaries, define two additional points, $i=0$ and $i=M+1$,
with $r_p(0) = 0$ and $f_p(M+1)=1$ for all $1\le p\le D$.  Define each
GRP by the set of $a_p=r_p(i_{pL})$ and $b_p=r_p(i_{pH})$, where the
$i_{pL}$ are the points which the low $r_p$ generalized faces of the GRP touch and
the $i_{pH}$ are the points the high $r_p$ generalized faces touch.  For $p=1$
$i_{1L}$ and $i_{1H}$ can be any pair of points with $0\le i_{1L} < i_{1H}\le
M+1$.  For $p>1$, $i_{pL}=0$ and/or $i_{pH}=M+1$ are allowable for a candidate
maximal GRP.  Other values of $i_{pL}$ and $i_{pH}$ for $p>1$ must for all
$1\le q < p$ satisfy the following criteria: 
$$r_p(i_{pL}) < r_p(i_{qL}),\ r_p(i_{qH}) < r_p(i_{pH}),$$
 and
$$r_q(i_{qL}) < r_q(i_{pL}),\ r_q(i_{pH}) < r_q(i_{qH}).$$
A simple way
to restrict the values of $i_{pL}$ and $i_{pH}$ that need to be checked
is to note that for $q=1$, this last condition is equivalent to
$$i_{1L} < i_{pL},\ i_{pH} < i_{1H}.$$
Once an acceptable GRP is found, One
may then go through all $1\le i \le M$ points and count the ones inside
the GRP, i.e., the ones which for all $1\le p\le D$ satisfy $r_p(i_{pL})
< r_p(i) < r_p(i_{pH})$.  But for $p=1$ that condition is equivalent to
 $i_{1L} < i < i_{1H}$, so only those values of $i$ need to be considered
as possibly being inside the GRP.  When the number of points inside the
GRP is $n$, the GRPs volume is a candidate for being the largest for
that value of $n$.  This algorithm can be applied to the data, and also
can be used with a Monte Carlo program which repeatedly generates sets of
events uniformly
in the unit generalized cube in order to compute the functions $C_{nD}(x;\mu)$.

A Monte Carlo program to evaluate $C_{\infty D}$ can begin with a unit
generalized cube divided into $N^D$ tiny generalized
sub-cubes, each with side of length $1/N$ for some large integer $N$.  For
each Monte Carlo trial, give every generalized sub-cube
an independent random value distributed according to a Gaussian
with zero mean and unit standard deviation.
Consider only GRPs defined by $a_p\le r_p \le b_p$ for all $1\le p \le D$
with the $\vec a=(a_1,...,a_p)$ and $\vec b=(b_1,...,b_p)$ on generalized sub-cube vertices.
A generalized volume $f$ GRP contains $N^Df$ generalized sub-cubes.  For
any such GRP, scale the contribution from all the
generalized sub-cubes it contains by a factor of $1/\sqrt{N^D f}$ to
produce a total signal $y$ with zero mean and unit standard deviation.
Consider all GRPs with $f$ within some bin for which $C_{\infty D}(y;f)$
is to be computed, with $\vec a$ on generalized sub-cube vertices, and
with the GRP entirely inside the unit generalized cube.  The optimum
GRP for the given $f$ bin is the one with smallest $y=y_{min}$.
Do a huge number of such Monte Carlo trials, and tabulate the
distribution of $y_{min}$ for the various bins of $f$. 

As for the $D=1$ case, there's a systematic upward shift in the computation
of $C_{\infty D}$ caused by use of finite $N$.
Correct for this systematic shift
by estimating what $y_{min}$ would be if the $\vec a$ of GRPs could
be shifted anywhere within each generalized sub-cube, instead of being
restricted to generalized sub-cube vertices.  The random minimum should have
the distribution expected for Gaussian Brownian motion in multiple ``time''
dimensions within the generalized sub-cube, given the GRP signal for $\vec a$
at each of the $2^D$ vertices of the generalized sub-cube.  Let's now discuss
how to choose a random signal with the correct distribution of the minimum
within each generalized cube.

Call ``$S_0$'' the GRP signal when $\vec a$ is at the vertex of the generalized
sub-cube with lowest $r_p$ for all $1\le p\le D$.  Call ``$S_k$''
the GRP signal when the GRP position is shifted to another generalized sub-cube
vertex for which only $r_k$ changes, leaving all other $r_p$ alone.  I.e,
the shift is parallel to axis $k$ of the unit generalized cube.  As the
GRP is continuously shifted, the signal continuously changes by adding
infinitesimal generalized volumes from one generalized face of the GRP and
removing infinitesimal generalized volumes from the opposite generalized
face.  Thus for such
shift the signal changes according to $S(t_k)=S_0 + \sigma_k w_k(t_k)$.  In
this equation, $t_k$ is $r_k$ shifted and rescaled so that it runs from 0 to 1
as $\vec a$ moves from one generalized sub-cube vertex to another along the
$k$ direction.  The $w_k(t_k)$ is a standard Wiener process, and $\sigma_k$ is
the total standard deviation of the signal change caused by a shift from
one generalized sub-cube vertex to its neighbor.
A total shift of $b_k-a_k$ would move the GRP exactly its total size.  It
would subtract a random signal with unit standard deviation from the original
GRP and add an independent random signal with unit standard deviation to the
shifted GRP.
The change in random signal would have variance equal to 2.  A shift by
a fraction of $b_k-a_k$ would change the signal by a random
value whose variance is the same fraction of 2.  Therefore a shift
by $1/N$ from one generalized sub-cube vertex to its neighbor along the $k$
direction changes the signal by a random value with variance
$$\sigma_k^2 = 2 \frac{1/N}{b_k-a_k}.$$ 
In the approximation that the generalized sub-cube is very small, a shift to
an arbitrary position within the generalized sub-cube gives a signal
$$S(\vec t) \approx S_0 + \sum_{k=1}^D \sigma_k w_k(t_k).$$
The $w_k$ functions are independent standard Wiener processes.  Each
one's minimum
can be chosen randomly according to the distribution of Eq. \ref{minw}
with $v=v_k=(S_k-S_0)/\sigma_k$.  So choose a set of independent random $P_k$,
each uniform over $(0,1)$; then choose
$z_k=\left( v_k-\sqrt{v_k^2 - 2\,ln(P_k)}\right)/2$ and take the minimum
signal within the generalized sub-cube to be $S_0 + \sum_{k=1}^D \sigma_k z_k$.

There are two remaining effects of the above described Monte Carlo procedure
which systematically shift $C_{\infty D}$ from its intended value for
$D>1$.  One such effect comes from binning in $f$, rather than using
exact values of $f$ as is possible with $D=1$.  As described, the Monte
Carlo program finds the smallest $y$ in a range of $f$, which will be
smaller than the smallest $y$ for a fixed value of $f$.  This downward
shift in $y$ gets smaller for smaller bins in $f$.  Another
source of a systematic shift in $C_{\infty D}$ is in the opposite direction.
The Monte Carlo program can only consider GRPs whose sides are all integer
multiples of $1/N$.  The true minimum of $y$ is almost surely less than
the minimum for such a restricted set of GRPs; so this limitation of the
method shifts $y$ up from its intended value by an amount which
gets smaller for larger $N$.  Neither of these two shifts would occur
if the GRPs were restricted to be generalized cubes, but such a restriction
would weaken the method's ability to avoid backgrounds which tend to be
concentrated near the end of the range of one of the $D$ variables while
being relatively uniform over the range of other variables.
One way of correcting for these two sources
of shift would be to generate tables using various bin sizes in $f$ and
various values of $N$; then extrapolate to zero bin size in $f$ and infinite
$N$.  Perhaps better methods can be found to improve the calculation
of $C_{\infty D}$.  But any such improvement is probably unnecessary.
Use of an approximate result need not have a
significant effect on computed upper limits.
If the same approximate function used in
place of $C_{\infty D}$ with the data is also used in the Monte Carlo
computation of $\bar C_{Max,D}(C,\mu)$, the result purporting to be
a $C$ confidence
level will still be a true $C$ confidence level, perhaps made conservative
by the presence of an unknown background.

\bibliographystyle{prsty}

\end{document}